\documentclass[aps,preprint]{revtex4}
\usepackage{psfig}

\begin{document}

\title{ \Large\bf $B\to f_{0}(980) K^{(*)}$ decays and final state
interactions }

\author{\large \bf Chuan-Hung Chen }


\affiliation{  Institute of Physics, Academia Sinica, Taipei,
Taiwan 115, Republic of China}

\date{\today}

\begin{abstract}
We study the exclusive decays of $B\to f_{0}(980) K^{(*)}$ in the
framework of the perturbative QCD by identifying the $f_{0}(980)$
as the composition of $\bar{s} s$ and
$\bar{n}n=(\bar{u}u+\bar{d}d)/\sqrt{2}$. We find that the
influence of the $\bar{n} n$ content on the predicted branching
ratios is crucial. We discuss the possible rescattering and
gluonium states which could enhance the branching ratios of
considered decays. We point out that the CP asymmetry in $B\to
f_{0}(980) K_{S,L}$ could be a new explorer of $\sin2\phi_{1}$.
\end{abstract}

\maketitle

Recently, Belle \cite{Belle1} and BaBar \cite{BaBar1} have opened
new channels in three-body nonleptonic $B$ decays, such as $B\to K
K K$, $K K\pi$ and $K \pi \pi$ decays.
In particular, BELLE has observed the decay of $B^{+}\to f_0(980)
K^{+}$ with the branching ratio (BR) product of $Br(B^{+}\to
f_{0}(980)K^{+})\times Br(f_{0}(980)\to \pi^+ \pi^-)=
(9.6^{+2.5+1.5+3.4}_{-2.3-1.5-0.8})\times 10^{-6}$ \cite{Belle1}.
Since $f_{0}(980)$ is a neutral scalar meson, the measured $B\to
f_{0}(980) K$ decays not only show that in the first time $B$
decays to scalar-pseudoscalar final states but also provide the
chance to understand the content of $f_0(980)$ and its production.

The essential inner structure of the scalar meson $f_{0}(980)$ is
still obscure since it was established first by Ref. \cite{f0}
with phase shift analysis. In the literature, $f_{0}(980)$ could
be four-quark states denoted by $qq\bar{q}\bar{q}$ \cite{4q} or
$K\bar{K}$ molecular states \cite{KK} or $\bar{q} q$ states
\cite{qq}. However, one objection against the possibility of
$K\bar{K}$ states is that the binding energy of $10-20$ MeV for
$K\bar{K}$ is much smaller than measured width in the range
$40-100$ MeV \cite{DP}. It is suggested that in terms of the
measured $\phi\to f_{0}(980) \gamma$ and $f_{0}(980)\to \gamma
\gamma$ \cite{DP,DLS,AAN,KVRS} decays and $D^{+}_{s}\to f_{0}(980)
\pi^{+}$ decay \cite {KVRS,dfk}, the flavor contents of
$f_{0}(980)$ are $\bar{s} {s}$ mostly and a small portion of
$\bar{n} n=(\bar{u}u+\bar{d} d)/\sqrt{2}$. In this paper, we take
$f_{0}(980)$ to be composed of $\bar{q} q$ states and use
$|f_{0}(980)>=\cos\phi_{s} |\bar{s} s>+\sin\phi_{s} |\bar{n} n>$
to denote its flavor wave function.

Before presenting our perturbative QCD (PQCD) calculation to the
decays, we would like to give a brief model-independent analysis
on $B\to f_{0}(980) K$.
For simplicity, our analysis will only concentrate on dominant
factorizable parts and regard the $f_{0}(980)$ as the composition
of $\bar{s} s$ so that at the quark level the process corresponds
to $b\to s\bar{s}s$ decay. Since at the quark level, the physics
for $B\to \phi K$ decays is the same as $B\to f_{0}(980)K^{(*)}$,
we first examine the $\phi K$ mode. We start by writing the
effective Hamiltonian for the $b\to s$ transition as
\begin{eqnarray*}
H_{{\rm eff}}=\frac{G_{F}}{\sqrt{2}}\sum_{q^{\prime
}=u,c}V_{q^{\prime
}}\left[ C_{1}(\mu ){\cal O}_{1}^{(q^{\prime })}+C_{2}(\mu ){\cal O}%
_{2}^{(q^{\prime })}+\sum_{i=3}^{10}C_{i}(\mu ){\cal
O}_{i}\right], \label{eff}
\end{eqnarray*}
where $V_{q^{\prime }}=V_{q^{\prime }s}^{*}V_{q^{\prime }b}$ are
the products of the CKM matrix elements, $C_{i}(\mu )$ are the
Wilson coefficients (WCs) and ${\cal O}_{i}$ correspond to the
four-quark operators. The explicit expressions of $C_{i}(\mu )$
and ${\cal O}_{i}$ can be found in Ref. \cite{BBL}. It is known
that the vector meson $\phi$ has the current matrix elements
\begin{eqnarray}
\langle 0|\bar{s} \gamma_{\mu} s| \phi\rangle=
m_{\phi}f_{\phi}\varepsilon_{\mu},\ \ \ \ \  \langle 0|\bar{s} s
|\phi\rangle=0. \label{mphi}
\end{eqnarray}
 Therefore, the decay amplitude of $B_d\to
\phi K^0$ can be simply described by \cite{Chen-PLB525}
\begin{eqnarray}
A\left( B_d\rightarrow \phi K^{0}\right) =  f_{\phi
}V_{t}^{*}\left( a_{3}^{(s)}+ a_{4}^{(s)} +a_{5}^{(s)}\right)
F_{e}^{BK} +2f_{B}V_{t}^{*}a_{6}^{(d)}F_{a6}^{\phi
K}+...\label{ampphik}
\end{eqnarray}
where $a^{(q)}_i$ are defined by
\begin{eqnarray}
a_{1}
&=&C_{1}+\frac{C_{2}}{N_{c}},\,a_{2}=C_{2}+\frac{C_{1}}{N_{c}} ,
\nonumber \\
a_{3,4}^{(q)} &=&C_{3,4}+\frac{3e_{q}}{2}C_{9,10}+a_{3,4}^{\prime
(q)},\,a_{3,4}^{\prime (q)}=\frac{C_{4,3}}{N_{c}}+\frac{3e_{q}}{2N_{c}}%
C_{10,9},  \nonumber \\
a_{5,6}^{(q)} &=&C_{5,6}+\frac{3e_{q}}{2}C_{7,8}+a_{5,6}^{\prime
(q)},\,a_{5,6}^{\prime (q)}=\frac{C_{6,5}}{N_{c}}+\frac{3e_{q}}{2N_{c}}%
C_{8,7}.  \label{ew}
\end{eqnarray}
Note that $a_{2}$ is larger than $a_{1}$ and $a^{(q)}_{4,6}$ are
much larger than $a^{(q)}_{3,5}$, $F_{e}^{BK}$ is the $B\to K$
decay form factor and $F_{a6}^{\phi K}$ denotes $\langle \phi
K|\bar{s} \gamma_{5} d|0\rangle$ annihilation effect. We neglect
$\langle \phi K|\bar{s} \gamma_{\mu} \gamma_{5} d|0\rangle$ due to
the chirality suppression. According to the results of Ref.
\cite{CKL-PRD}, the predicted BR of $B_{d} \to \phi K^0$ is around
$\sim 9.6\times 10^{-6}$ which is consistent with Belle's and
BaBar's results given by $(10.0^{+1.9+0.9}_{-1.7-1.3})\times
10^{-6}$ \cite{Belle2} and $(8.7^{+1.7}_{-1.5}\pm 0.9)\times
10^{-6}$ \cite{BaBar2}, respectively. However, in the $B\to
f_{0}(980) K^0 $ decay the relevant current matrix elements are
given by
\begin{eqnarray}
\langle 0|\bar{s} \gamma_{\mu} s| f_{0}(980)\rangle = 0; \ \ \ \ \
\langle 0|\bar{s} s |f_{0}(980)\rangle = m_{f_{0}} \tilde{f}
\label{mf}
\end{eqnarray}
where $m_{f_{0}}(\tilde{f})$ are the mass ( decay constant) of
$f_{0}(980)$. From Eqs. (\ref{mphi}) and (\ref{mf}), it is clear
that the situation in the $f_0(980) K^0$ mode is just opposite to
the case of $\phi K^0$, i.e., the role in vector and scalar vertex
is exchanged each other. Hence, the decay amplitude for $B\to
f_{0}(980) K^0$  can be simply written as
\begin{eqnarray}
A\left( B_d\rightarrow f_{0}(980) K^{0}\right) = 2r_{f} \tilde{f}
V_{t}^{*} a_{6}^{(s)} S_{e}^{BK}
+2f_{B}V_{t}^{*}a_{6}^{(d)}F_{a6}^{f_{0} K}+...\label{ampfk}
\end{eqnarray}
where $r_{f}=m_{f_{0}}/M_{B}$, $S_{e}^{BK}=\langle K|\bar{b}
s|B\rangle/M_{B}$ and the factor of $2$ comes from Fierz
transformation in the $(V-A)\times (V+A)$ operator. By taking
$M_{B} \approx m_{b}$ with $m_{b}$ being the $b$-quark mass, one
can expect $S_{e}^{BK}\approx F_{e}^{BK}$ by equation of motion.
Comparing Eq. (\ref{ampphik}) with Eq. (\ref{ampfk}), it is
obvious that there is a suppression factor $r_{f}=0.186$ in $B\to
f_{0}(980) K^{0}$. Including the factor of 2, taking
$a^{(s)}_{6}(\sqrt{\bar{\Lambda}M_{B}})/a^{(s)}_{4}(\sqrt{\bar{\Lambda}M_{B}})\approx
1.5 $ and $\tilde{f}\sim f_{\phi}$ and neglecting the annihilation
contributions, we estimate the ratio of $Br(B\to f_{0}(980)
K^0)/Br(B\to \phi K^0)$ being around $0.31$. With the average
value of Belle and BarBar, the predicted $Br(B\to f_{0}(980) K^0)$
is around $2.8 \times 10^{-6}$. Will the value change while we
include the $\bar{n} n$ content and annihilation contributions? In
the following we display the results in a more serious theoretical
approach.

It is known that large uncertainities are always involved in the
calculations of transition matrix elements while studying
exclusive hadron decays. Nevertheless, the problem will become
mildly in the heavy $B$ meson decays because of the enormous $B$
samples produced by $B$ factories. That is, more precise
measurements will help the theory to pin down the unknown
parameters to make some predictions.
In this paper, we adopt the PQCD approach in which the
applications to exclusive heavy $B$ meson decays, such as
$B\rightarrow K \pi$ \cite{KLS}, $B\rightarrow \pi \pi(KK) $
\cite{LUY,CL-PRD,Chen-PLB520}, $B\rightarrow \phi \pi(K)$
\cite{Melic,CKL-PRD}, $B\rightarrow \eta^{(')} K$ \cite{KS} and
$B\rightarrow \rho K$ \cite{Chen-PLB525} decays, have been studied
and found that all of them are consistent with the current
experimental data \cite{SU,Keum}. In the PQCD, in order to solve
the various divergences encountered at end-point, we will include
not only $k_{T}$ resummation, for removing end-point
singularities, but also threshold resummation, for smearing the
double logarithmic divergence arisen from weak corrections
\cite{KLS-PRD}.

In order to satisfy the local current matrix elements with Eq.
(\ref{mf}), the $f_0(980)$ meson distribution amplitude is given
by
\begin{eqnarray}
\langle 0|\bar{q}(0)_{j}
q(z)_{l}|f_{0}\rangle &=&\frac{1}{\sqrt{2N_{c}}}\int^{1}_{0} dx
e^{-ixP\cdot z} m_{f}[{\bf 1}]_{lj} \Phi_{f_{0}}(x) \label{daf}
\end{eqnarray}
with the normalization
\begin{eqnarray*}
\int^{1}_{0} dx\Phi_{f_0}(x)=\frac{\tilde{f}}{2\sqrt{2N_{c}}}.
\end{eqnarray*}
For $B$ and $K^{(*)}$ mesons, the corresponding distribution
amplitudes \cite{CKL-PRD,KLS-PRD} are written as
\begin{eqnarray}
\langle 0|\bar{b}(z)_{j}q(0)_{l}|B\rangle &=&
\frac{1}{\sqrt{2N_{c}}}\int^{1}_{0} dx e^{-izxP\cdot x}
([\not{P}]_{jl}+M_{B}[I]_{jl})\gamma_{5}\phi_{B}(x)\\
\langle K|\bar{q}(z)_{j}s(0)_{l}|0\rangle
&=&\frac{1}{\sqrt{2N_{c}}}\int^{1}_{0} dx e^{-izP\cdot x}
\Big\{[\gamma_{5}\not{P}]_{jl} \Phi_{K}(x) \nonumber \\
&& +m^{0}_{K} [\gamma_{5}]_{jl} \Phi^{p}_{K}(x) +m^{0}_{K}
[\gamma_{5}(\not{n}_{-}\not{n}_{+}-1)]_{jl}
\Phi^{\sigma}_{K}(x)\Big\}, \label{dak}\\
\langle K^*(\varepsilon_{L})|\bar{q}(z)_{j}s(0)_{l}|0\rangle &=&
\frac{1}{\sqrt{2N_{c}}}\int^{1}_{0} dx e^{-izP\cdot x}
\Big\{ M_{K^*}[\not{\varepsilon}_{L}]_{lj} \phi_{K^*}(x) \nonumber \\
&& +[\not{\varepsilon}_{L}\not{P}]_{lj}
\phi^{t}_{K^*}(x)+M_{K^*}[I]_{lj} \phi^{s}_{K^*}(x)\Big\}
\label{daks}
\end{eqnarray}
with $n_{-}=(0,1,0_{\perp})$, $n_{+}=(1,0,0_{\perp})$ and
$m^{0}_{K}$ being the chiral symmetry breaking parameter,
$\Phi_{K^{(*)}}(x)$ stand for twist-2 wave functions and the
remains belong to twist-3 with their explicit expressions being
given in Ref. \cite{CKL-PRD,BBKT}. Although $K^*$ meson has three
polarizations, only the longitudinal part is involved in the
decays of $B\to f_{0}(980) K^{*}$. We note that as the $K$ meson,
the distribution amplitude of $f_{0}(980)$ can have more
complicated spin structures. However, since our purpose is just
for the properties of $B\to f_{0}(980) K^{(*)}$, we consider only
the simplest case. Moreover, if $f_{0}(980)$ consists of $\bar{s}
s$ mostly, the choice of Eq. (\ref{daf}) is clearly dominant .

The decay rates of $B\to f_{0}(980) K$ are expressed by
\begin{equation}
\Gamma =\frac{G_{F}^{2}M_{B}^{3}}{32\pi }|A|^{2}
\end{equation}
where $A$ includes all possible components of $f_{0}(980)$ and
topologies. As mentioned before, $f_{0}(980)$ has the components
of $\bar{s} s$ and $\bar{n} n$, for different contents, the
amplitudes of $B_{d}\rightarrow f_{0}(980) \bar{K}^{0}$ and
$B^{+}\rightarrow f_{0}(980)  K^{+}$ are written as
\begin{eqnarray}
A_{\bar{s} s} &=& \tilde{f} V_{t}^{*}S_{e6}^{P(s)}
+f_{B}V_{t}^{*}S_{a46}^{P(d)}+...,
\nonumber \\
A_{\bar{n} n} &=&f_{K }V_{t}^{*}N_{e46}^{P(d)}
+f_{B}V_{t}^{*}N_{a46}^{P(d)}+...,
\nonumber \\
A^{+}_{\bar{s}s} &=&\tilde{f} V_{t}^{*}
S_{e6}^{P(s)}+f_{B}V_{t}^{*}S_{a46}^{P(u)} -f_{B}V_{u}^{*}S_{a}+...,\nonumber \\
A^{+}_{\bar{n}n} &=&f_{K} V_{t}^{*}
N_{e46}^{P(u)}+f_{B}V_{t}^{*}N_{a46}^{P(u)}-f_{K}V_{u}^{*}N_{e}
-f_{B}V_{u}^{*}N_{a}+...,\label{amps}
\end{eqnarray}
respectively, where $S^{P(q)}_{e(a)}\ (N^{P(q)}_{e(a)})$ denote
the emission (annihilation) contributions of the $\bar{s} s \
(\bar{n} n)$ content from penguin diagrams while $S_{e(a)}
(N_{e(a)})$ are from tree contributions. The total decay amplitude
for the neutral (charged) mode is described by $A=\cos\phi
A^{(+)}_{\bar{s} s} +\sin\phi A^{(+)}_{\bar{n} n}/\sqrt{2}$. Due
to the smallness of nonfactorizable effects, we neglect to show
them in the Eq. (\ref{amps}) and just display the factorizable
contributions with emission and annihilation topologies. But we
will include their effects in our final numerical results.
According to Eqs. (\ref{daf}$-$\ref{dak}), the hard amplitudes for
$\bar{s} s$ content are derived as
\begin{eqnarray}
S^{P(q)}_{e6}&=&16\pi r_{f} \alpha_{s} C_{F} M^{2}_{B}
\int_{0}^{1}dx_{1}dx_{3}\int_{0}^{\infty}
b_{1}db_{1}b_{3}db_{3}\Phi _{B}( x_{1},b_{1})
\nonumber\\
&& \times \Big\{ \Big [\Phi_{K}(x_{3}) +2
r_{K}\Phi^{p}_{K}(x_{3})+r_{K}x_{3}(\Phi^{p}_{K}(x_3)-\Phi^{\sigma}_{K}(x_{3}))\Big]
\nonumber\\
&& \times ES^{(q)}_{e6}(t^{(1)}_{e})
h_{e}(x_{1},x_{3},b_{1},b_{3})
\nonumber\\
&&+2r_{K}\Phi^{p}_{K}(x_3)ES^{(q)}_{e6}(t^{(2)}_{e})
h_{e}(x_{3},x_{1},b_{3},b_{1}) \Big\},
\end{eqnarray}
\begin{eqnarray}
S_{a4}^{P( q) } &=&-16\pi r_{f}r_{K}
C_{F}M_{B}^{2}\int_{0}^{1}dx_{2}dx_{3}\int_{0}^{\infty
}b_{2}db_{2}b_{3}db_{3}\Phi _{f_{0}}( x_{2})
\nonumber \\
&&\times \bigg\{ \Big[ (1+x_{3})\Phi_{K}^{p}(1- x_{3})
+(1-x_{3})\Phi_{K}^{\sigma }( 1-x_{3}) \Big] E_{a4}^{(q)}(
t_{a}^{( 1) }) h_{a}( x_{2},x_{3},b_{2},b_{3})
\nonumber \\
&&-\Big[ (1+x_{2}) \Phi_{K}^{p}( 1-x_{3}) \Big] E_{a4}^{(q)}(
t_{a}^{( 2) }) h_{a}(x_{3},x_{2},b_{3},b_{2}) \bigg\}, \label{Sa4}
\end{eqnarray}
\begin{eqnarray}
 S_{a6}^{P( q) } &=&16\pi r_{f}
C_{F}M_{B}^{2}\int_{0}^{1}dx_{2}dx_{3}\int_{0}^{\infty}
b_{2}db_{2}b_{3}db_{3} \Phi_{f_{0}}(x_{2})\Phi_{K} (1- x_{3})
\nonumber \\
&&\times \bigg\{2 E_{a6}^{(q)}( t_{a}^{( 1) }) h_{a}(
x_{2},x_{3},b_{2},b_{3}) + x_{2} E_{a6}^{(q)}( t_{a}^{( 2) })
h_{a}(x_{3},x_{2},b_{3},b_{2}) \bigg\} \label{Sa6}
\end{eqnarray}
and the results for the content of $\bar{n} n$ are given by
\begin{eqnarray}
N^{P(q)}_{e4}&=&8\pi r_{f} \alpha_{s} C_{F} M^{2}_{B}
\int_{0}^{1}dx_{1}dx_{2}\int_{0}^{\infty}
b_{1}db_{1}b_{2}db_{2}\Phi _{B}( x_{1},b_{1})\Phi_{f_{0}}(x_{2})
\nonumber\\
&& \times \Big\{ (1-2x_{2})
 EN^{(q)}_{e4}(t^{(1)}_{e}) h_{e}(x_{1},x_{2},b_{1},b_{2})
+2EN^{(q)}_{e4}(t^{(2)}_{e})
h_{e}(x_{2},x_{1},b_{2},b_{1}) \Big\} \label{Ne4}\\
N^{P(q)}_{e6}&=&-16\pi r_{f} r_{K} \alpha_{s} C_{F} M^{2}_{B}
\int_{0}^{1}dx_{1}dx_{2}\int_{0}^{\infty}
b_{1}db_{1}b_{2}db_{2}\Phi _{B}( x_{1},b_{1})\Phi_{f_{0}}(x_{2}),
\nonumber\\
&& \times \Big\{ (2+x_{2})
 EN^{(q)}_{e6}(t^{(1)}_{e}) h_{e}(x_{1},x_{2},b_{1},b_{2})
+2EN^{(q)}_{e6}(t^{(2)}_{e}) h_{e}(x_{2},x_{1},b_{2},b_{1}) \Big\}
.\label{Ne6}
\end{eqnarray}
Here, the hard part functions $h_{e(a)}$, mainly arising from the
propagators of hard gluon and valence quark, have included the
threshold resummation factor \cite {CKL-PRD}, and the evolution
factors are given by
\begin{eqnarray*}
ES_{e6}^{(q)}(t) &=&\alpha _{s}(t)a_{6}^{(q)}(t)\exp
[-S_{B}(t,x_{1})-S_{K}(t,x_{3})], \\
EN_{ei}^{(q)}(t) &=&\alpha _{s}(t)a_{i}^{(q)}(t)\exp
[-S_{B}(t,x_{1})-S_{f_{0}}(t,x_{2})], \\
E_{ai}^{(q)}(t) &=&\alpha _{s}(t)a_{i}^{(q)}(t)\exp
[-S_{f_{0}}(t,x_{2})-S_{K}(t,x_{3})]
\end{eqnarray*}
where the exponential effects denote Sudakov factors generated by
the $k_{T}$ resummation. Since the annihilation topologies in both
contents are associated with the matrix elements $<f_{0}(980)
K|\bar{s}\gamma_{\mu}\gamma_{5}d(u)|0 >$ and $<f_{0}(980)
K|\bar{s}\gamma_{5} d(u)|0 >$, the difference in the different
content is only from the spectator, which is $s\ (\texttt{{d or
u}})$ in $\bar{s} s\ (\bar{n} n)$ component. Hence,
$N^{P(q)}_{a4(6)}$ could be obtained from $S^{P(q)}_{a4(6)}$ by
replacing $x_{2}$ with $1-x_{2}$ in $\Phi_{f_{0}}(x_{2})$ and
$1-x_{3}$ with $x_{3}$ in $\{\Phi_{K}(1-x_{3})\}$. In Eq.
(\ref{amps}), we define that
$S^{P(q)}_{a46}=S^{P(q)}_{a4}+S^{P(q)}_{a6}$ and
$N^{P(q)}_{e46}=N^{P(q)}_{e4}+ N^{P(q)}_{e6}$. The $S_{a}$ and
$N_{e}$ can be obtained from $S^{P(q)}_{a4}$ and $N^{P(q)}_{e4}$
by replacing W.C. $a^{(q)}_{4}$ with $a_{2}$. We note that unlike
the cases of $B\to PP$ and $B\to PV$ decays in which the chirality
suppression in the $(V-A)\times (V-A)$ annihilation topology is
conspicuous, the contribution of Eq. (\ref{Sa4}) may not be small
because one twist-3 effect, such as $\Phi^{\sigma}_{K}$, is not
cancelled. And also, differing from $B\to PP$ decays, Eq.
(\ref{Ne4}) is opposite to Eq. (\ref{Ne6}) in sign.

Similar to $B\to f_{0}(980) K$ decays, the amplitudes for $B\to
f_{0}(980) K^{*0}$ and $B^{+}\to f_{0}(980) K^{*+}$  can be
expressed as
\begin{eqnarray}
A_{\bar{s} s} &=& \tilde{f} V_{t}^{*}K_{e6}^{P(s)}
+f_{B}V_{t}^{*}K_{a46}^{P(d)}+...,
\nonumber \\
A_{\bar{n} n} &=&f_{K^* }V_{t}^{*}X_{e4}^{P(d)}
+f_{B}V_{t}^{*}X_{a46}^{P(d)}+...,
\nonumber \\
A^{+}_{\bar{s}s} &=&\tilde{f} V_{t}^{*}
K_{e6}^{P(s)}+f_{B}V_{t}^{*}K_{a46}^{P(u)} -f_{B}V_{u}^{*}K_{a}+...,\nonumber \\
A^{+}_{\bar{n}n} &=&f_{K^*} V_{t}^{*}
X_{e4}^{P(u)}+f_{B}V_{t}^{*}X_{a46}^{P(u)}-f_{K^*}V_{u}^{*}X_{e}
-f_{B}V_{u}^{*}X_{a}+....\label{ampks}
\end{eqnarray}
where $K^{P(q)}_{e6}$,  $K^{P(q)}_{a4}$ and $K^{P(q)}_{a6}$ can be
obtained from $S^{P(q)}_{e6}$, $S^{P(q)}_{a4}$ and
$S^{P(q)}_{a6}$, respectively, by replacing $\phi_{K}$,
$\phi_{K}^{p}$ and $\phi_{K}^{\sigma}$ with $\phi_{K^*}$,
$\phi_{K^*}^{s}$ and $\phi_{K^{*}}^{t}$. With the similar
analysis, we find that $X_{e4}=-N_{e4}$ and $X_{a4(6)}^{P(q)}$ are
related to $K_{a4(6)}^{P(q)}$ if we change $x_{2}$ and $1-x_{3}$
to $1-x_{2}$ and $x_{3}$, respectively. The $K_{a}$ and $X_{e}$
are the same as $K^{P(q)}_{a4}$ and $X^{P(q)}_{e4}$ but the
associated WC is $a_{2}$.

\begin{figure}[bhpt]
 \centerline{ \psfig{figure=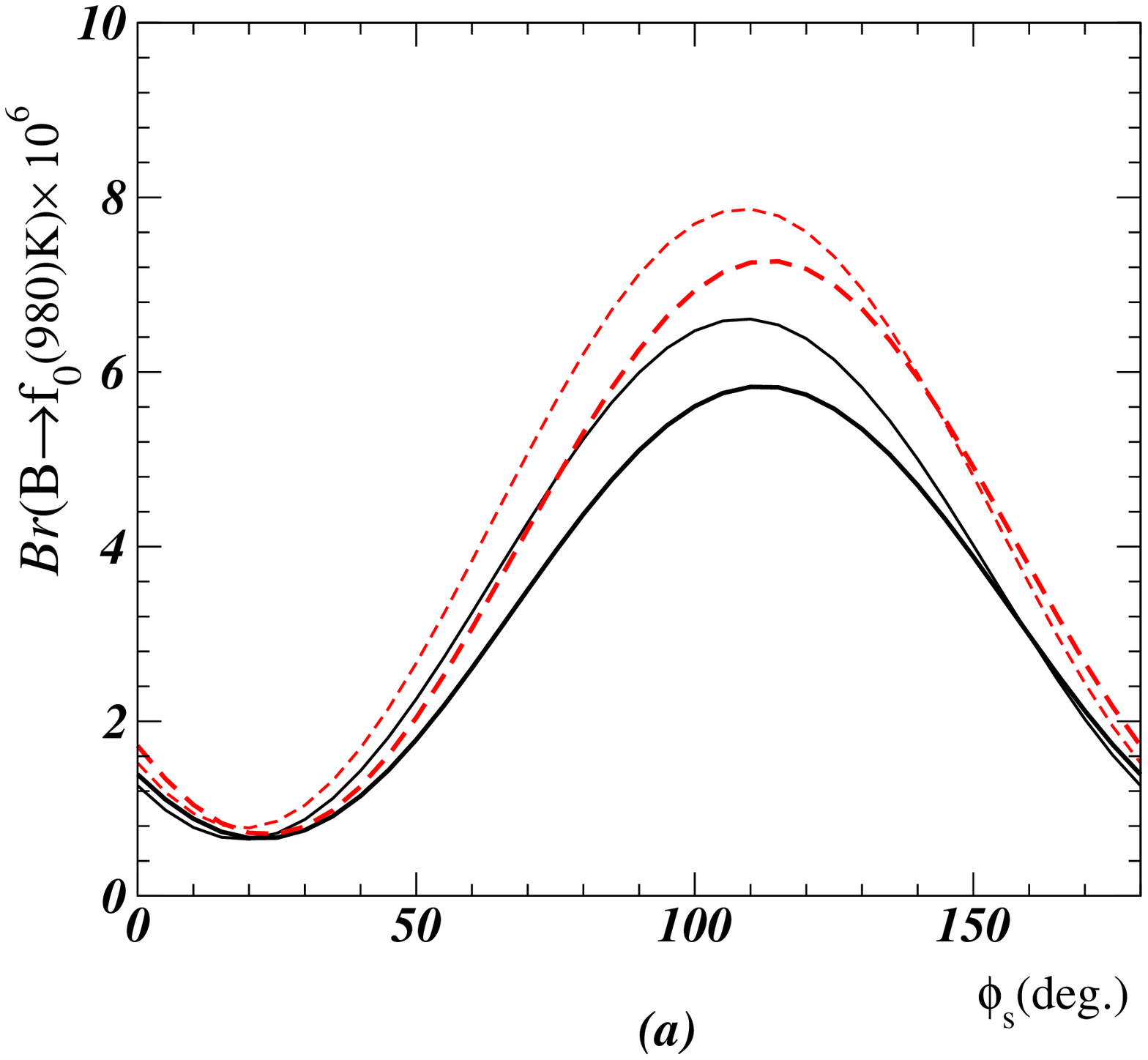,height=2.2in }
\psfig{figure=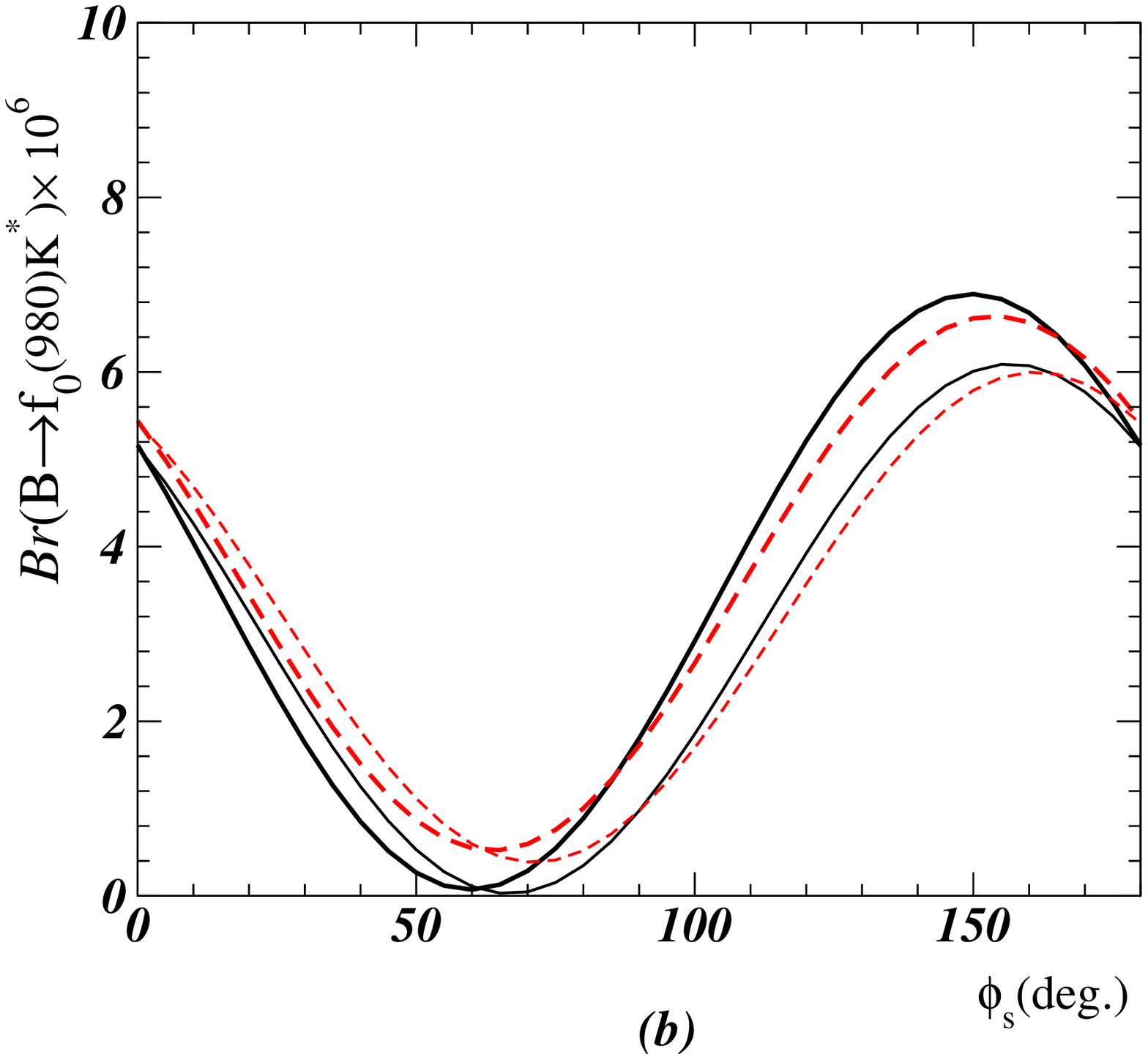,height=2.2in}} \caption{BRs of (a) $B\to
f_{0}(980) K$ and (b) $B\to f_{0}(980) K^{(*)}$ as a function of
$\phi_{s}$. The bold (thin) solid lines stand for neutral modes
with $\xi=0.3(0.5)$ while the dashed lines are for charged modes.}
\label{figbr}
\end{figure}

 In our numerical calculations, the $B$ meson wave function is
taken as \cite{CL-PRD}
\begin{eqnarray*}
\Phi _{B}(x,b)&=&N_{B}x^{2}(1-x)^{2}\exp \left[ -\frac{1}{2}\left( \frac{xM_{B}%
}{\omega _{B}}\right) ^{2}-\frac{\omega _{B}^{2}b^{2}}{2}\right]
\label{bwf}
\end{eqnarray*}
where $\omega_{B}$ is the shape parameter and $N_{B}$ is the
normalization, determined by
\begin{eqnarray*}
\int^{1}_{0} dx \Phi_{B}(x,b=0)=\frac{f_{B}}{2\sqrt{2N_{c}}}.
\end{eqnarray*}
 Since $f_{0}(980)$ is a light meson, its wave function can be
defined in the frame of the light-cone and the concept of the
twist expansion can be used. And because the relevant scalar meson
wave function has not been derived in the literature yet, we
choose the following form
\begin{eqnarray*}
\Phi_{f_0}(x)&=&\frac{\tilde{f}}{2\sqrt{2N_{c}}} \bigg\{
3(1-2x)^{2}+ \xi (1-2x)^{2}(C^{3/2}_{2}(1-2x)-3)+1.8 C^{1/2}_4(1-2x) \bigg\}
 \label{phi3t}
\end{eqnarray*}
with $C^{1/2}_{4}(y)=(35y^4 -30 y^2+3)/8$,
$C^{3/2}_{2}(y)=3/2(5y^2-1)$ and $\xi=0.3\sim 0.5$ in our
estimations.
In order to fix the values of $\omega_{B}$ and $m^{0}_{K}$, we
take $\omega_{B}=0.4$ and $m^{0}_{K}=1.7$ GeV as those in the
studies of the $B\to K$ and $B\to K^*$ form factors
\cite{CL-PRD,CQ}. Explicitly, with the values above and
 $\tilde{f}=0.2$ \cite{DP}, the $B\to f_{0}(980)$ form factor is
found to be 0.270 (0.286) for $\xi=0.3\ (0.5)$. Using Eqs.
(\ref{amps}$-$\ref{ampks}), the values of hard amplitudes are
shown in Table \ref{tabamp}.
\begin{table}[hbt]
\caption{The parameters in the amplitudes with $\omega_{B}=0.4$,
$m^{0}_{K}=1.7$ and $\tilde{f}=0.2$ GeV for $\xi=0.3$.
}\label{tabamp}
\begin{center}
\begin{tabular}{c|ccccc}
\hline
Amp. &  $ \ S^{P(s)}_{e6}(10^{-2}) \  $ & $  \ S^{P(d)}_{a46}(10^{-2}) \  $
& $ \  S_{a}(10^{-2}) $ & $ N^{P(u)}_{e46}(10^{-2}) \ $ &
$ \ N_{e}(10^{-2}) \ $ \\
\hline $f_{0}(980)K$ & $-1.02$  &  $0.45+i0.39$ &  $-7.41-i0.40$ &
$1.75 $ & $31.44$ \\ \hline Amp. &  $ \ K^{P(s)}_{e6}(10^{-2}) \ $
& $  \ K^{P(d)}_{a46}(10^{-2}) \  $ & $ \  K_{a}(10^{-2}) $ & $
X^{P(u)}_{e4}(10^{-2}) \ $ & $ \ X_{e}(10^{-2}) $ \\ \cline{1-6}
$f_{0}(980)K^*$ & $-1.37$ & $0.06+i0.40$ & $0.83-i1.98$ & $1.31$ &
$-31.44$\\ \hline
\end{tabular}
\end{center}
\end{table}
In addition, the decay BRs and CP asymmetries (${\cal A}_{CP}$)
for $B\to f_{0}(980) K$ and $B\to f_{0}(980) K^*$ are shown in
Table \ref{xi} with $\phi_{s}=0$ and in Table \ref{phis} with
$\phi_{s}=40^{0}$ and $140^{0}$, where
\begin{eqnarray*}
{\cal A}_{CP}={\bar{\Gamma} -\Gamma\over \bar{\Gamma} + \Gamma}
\end{eqnarray*}
with $\Gamma\ (\bar{\Gamma})$ being the particle (antiparticle)
partial decay rate. In both tables, the CP violating phase of
$V_{ub}$ is taken to be $\phi_{3}=72^{0}$. From Table
\ref{tabamp}, we see clearly that the sign of annihilation
contribution $S^{P(d)}_{a46}$ is opposite to $S^{P(s)}_{e6}$ so
that the BRs of $B\to f_{0}(980) K$ are reduced significantly.
Because the corresponding value of $S^{P(d)}_{a46}$ in $B\to
f_{0}(980) K^{*}$ is too small as shown in Table \ref{xi}, we can
understand that the influence of the annihilation is
insignificant. Also we see that if $f_{0}(980)$ is $\bar{s} s$
mostly,  BRs of $B\to f_{0}(980) K$ are only around $10^{-6}$. On
the contrary, they could be over $5.0\times 10^{-6}$ at a large
angle of $\phi_{s}$; whereas BRs for $B\to f_{0}(980) K^{*}$
 are similar to those at $\phi_{s}=0$. To further understand the
effect of $\bar{n} n$, we display the BRs as a function of
$\phi_{s}$ in Figure \ref{figbr}. Obviously, the predictions are
very sensitive to the contributions of the $\bar{n} n$ component.
The essential question is what the rang of $\phi_{s}$ is allowed.
Unfortunately, the preferable $\phi_{s}$ is still unknown, {\it
i.e.}, both small \cite{Ochs} and large \cite{AAN} angle solutions
exist simultaneously in the literature. The former prefers the
value of $\phi_{s}=42.14^{+5.8^{0}}_{-7.3}$ but the latter
$138^{0}\pm {6^0}$.

It is also interesting to look for a decay which is directly
related to the $\bar{n} n$ content only such that it can tell us
how the impact of the content would be. We point out that  one of
the possible decays is $B^{+}\to f_{0}(980) \pi^{+}$. In this
decay, from the identity of Eq. (\ref{mf}) the contribution from
the $\bar{s} s$ content vanishes since it corresponds to the
vector current vertex. As the $\pi$ meson is similar to $K$ except
$SU(3)$ breaking effects as well as the relevant wave functions
and CKM matrix elements, the BR is estimated to be $0.25 \ (0.45)
\times 10^{-6}$ for $\phi_{s}=35^{0}\ (132^{0})$.

\begin{table}[hbt]
\caption{BRs of $B\to f_{0}(980) K^{(*)}$ without and with
annihilation contributions by taking the same values of parameters
as Table \ref{tabamp}. The CPAs with annihilation effects are also
shown. }\label{xi}
\begin{center}
\begin{tabular}{c|ccc}
\hline 
Mode &  \begin{tabular}{c} BR($\times 10^{-6}$)\\
no anni.
\end{tabular} & BR($\times 10^{-6}$) & CPA ($\%$) \\
\hline $B\to f_{0}(980) K^0$
& $\hspace{0.7cm} $2.95$ \hspace{0.7cm}$  & $\hspace{0.7cm} $1.39$
\hspace{0.7cm}$ &
$\hspace{0.7cm} 0 \hspace{0.7cm}$\\
\hline $B^+\to f_{0}(980) K^+$ &
 $\hspace{0.7cm} 3.13 \hspace{0.7cm}$ &
$\hspace{0.5cm} 1.57 \hspace{0.5cm}$ &
$\hspace{0.5cm} 6.50 \hspace{0.5cm}$\\
\hline
$B\to f_{0}(980) K^{*0}$ &
$\hspace{0.7cm} 5.16 \hspace{0.7cm}$ & $\hspace{0.7cm} 5.40
\hspace{0.7cm}$ &
$\hspace{0.7cm} 0 \hspace{0.7cm}$\\
\hline $B^+\to f_{0}(980) K^{*+}$ &
 $\hspace{0.7cm} 5.49 \hspace{0.7cm}$ &
$\hspace{0.5cm} 5.76 \hspace{0.5cm}$ & $\hspace{0.5cm} 1.48
\hspace{0.5cm}$ \\ \hline
\end{tabular}
\end{center}
\end{table}

\begin{table}[hbt]
\caption{BRs and CPAs of $B\to f_{0}(980) K^{(*)}$ decays with the
same parameters as Table \ref{tabamp} but by taking
$\phi_{s}=40^0$ and $140^0$ . }\label{phis}
\begin{center}
\begin{tabular}{c|ccc}
\hline
Mode & $\phi_{s}$ & BR\Big($\times 10^{-6}\Big)$ & CPA \Big($\%$\Big) \\
\hline $B\to f_{0}(980) K^0$ & $\hspace{0.7cm} 40^{0}
\hspace{0.7cm}$ & $\hspace{0.7cm} 1.14 \hspace{0.7cm}$ &
$\hspace{0.7cm} 0 \hspace{0.7cm}$\\
\cline{2-4}  
&$\hspace{0.5cm} 140^{0} \hspace{0.5cm}$& $\hspace{0.5cm} 4.70 \hspace{0.5cm}$ &
$\hspace{0.5cm} 0 \hspace{0.5cm}$ \\
\hline $B^+\to f_{0}(980) K^+$ & $\hspace{0.5cm} 40^{0}
\hspace{0.5cm}$ & $\hspace{0.5cm} 1.25 \hspace{0.5cm}$ &
$\hspace{0.5cm} -16.94 \hspace{0.5cm}$\\
\cline{2-4} 
&$\hspace{0.5cm} 140^{0} \hspace{0.5cm}$ &$\hspace{0.5cm}
5.94 \hspace{0.5cm}$& $\hspace{0.5cm} 1.58 \hspace{0.5cm}$ \\
\hline
$B\to f_{0}(980) K^{*0}$ & $\hspace{0.7cm} 40^{0} \hspace{0.7cm}$
& $\hspace{0.7cm}  0.85 \hspace{0.7cm}$ &
$\hspace{0.7cm} 0 \hspace{0.7cm}$\\
\cline{2-4} 
&$\hspace{0.5cm} 140^{0} \hspace{0.5cm}$& $\hspace{0.5cm}
6.70 \hspace{0.5cm}$ &
$\hspace{0.5cm} 0 \hspace{0.5cm}$ \\
\hline $B^+\to f_{0}(980) K^{*+}$ & $\hspace{0.5cm} 40^{0}
\hspace{0.5cm}$ & $\hspace{0.5cm} 1.51 \hspace{0.5cm}$ &
$\hspace{0.5cm} 11.31 \hspace{0.5cm}$\\
\cline{2-4} 
 &$\hspace{0.5cm} 140^{0} \hspace{0.5cm}$ &$\hspace{0.5cm}  6.30
\hspace{0.5cm}$& $\hspace{0.5cm}  -12.19 \hspace{0.5cm}$ \\\hline
\end{tabular}
\end{center}
\end{table}

\begin{figure}[bhpt]
 \centerline{ \psfig{figure=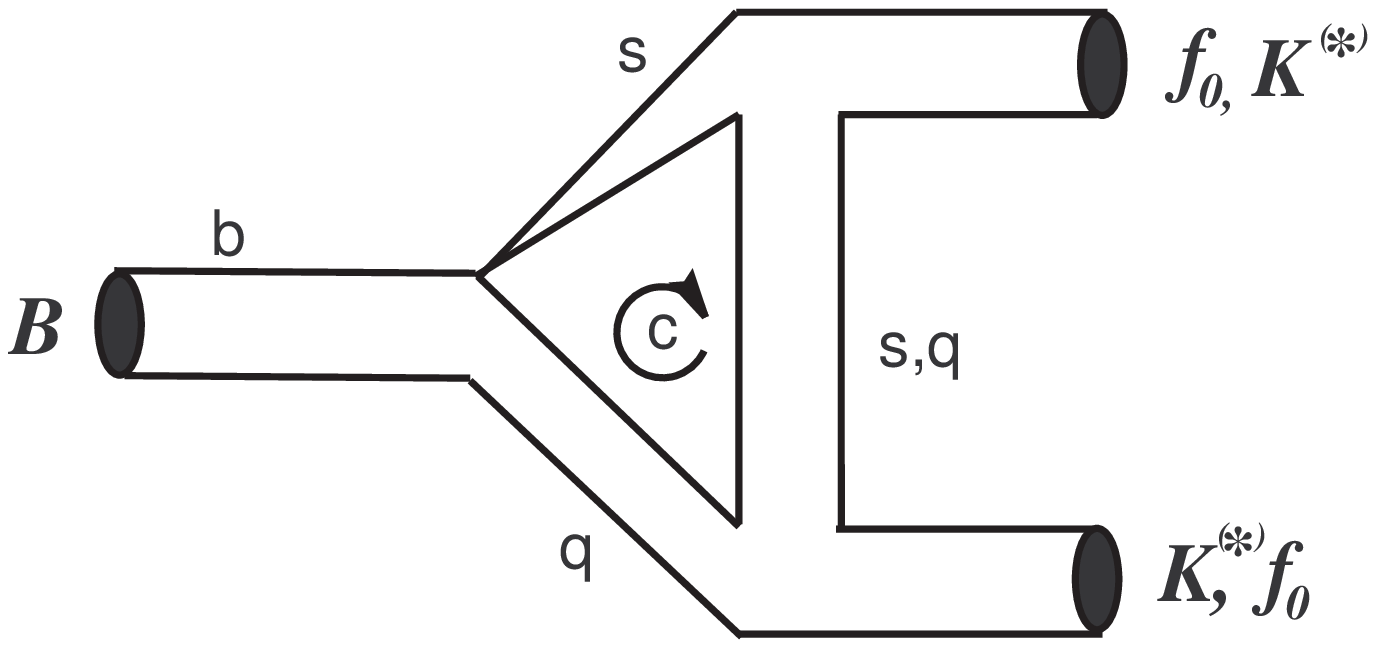,height=1.3in }
\psfig{figure=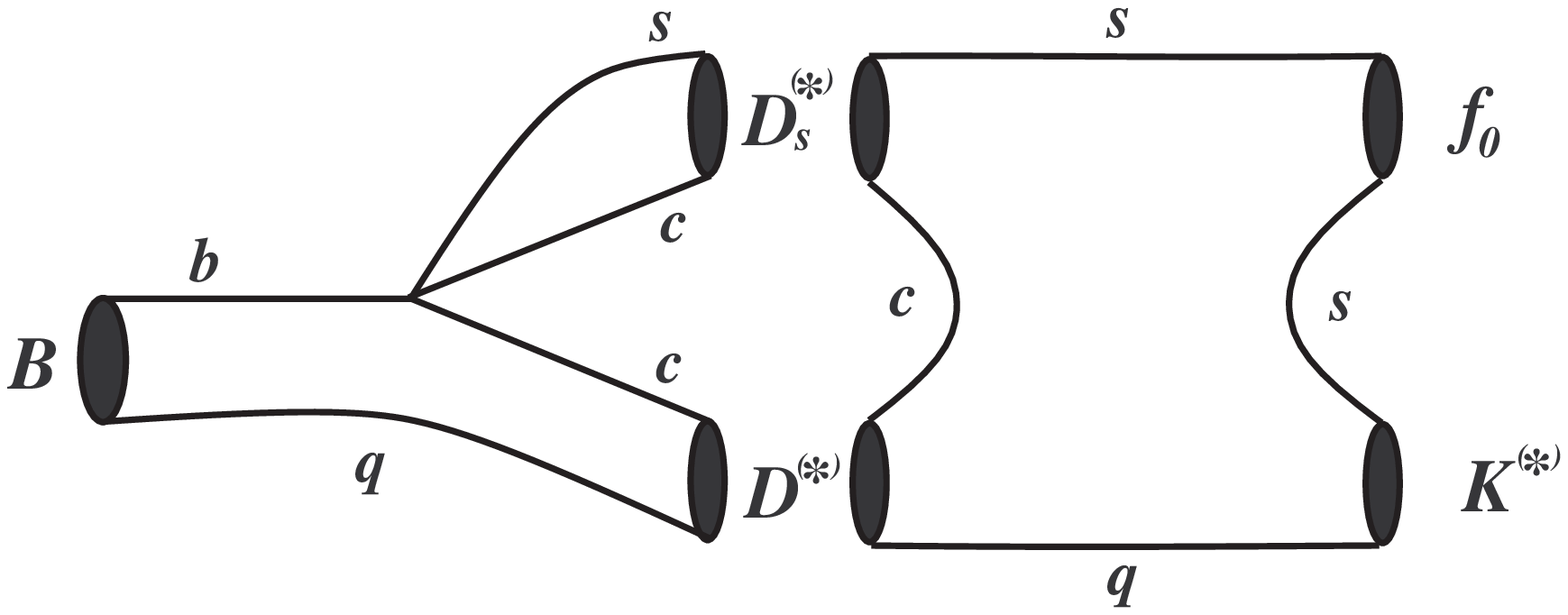,height=1.3in}} \caption{Possible
rescattering diagrams contributing to $B\to f_{0}(980)
K^{(*)}$.}\label{res}
\end{figure}

It seems that in spite of the uncertain allowed value of
$\phi_{s}$, the predicted BR of $B^{+}\to f_{0}(980) K^{+}$ is
smaller than the central (minimal) value of $21.17 \ (11.03)\times
10^{-6}$ reported by Belle in which $Br(f_{0}(980)\to \pi^+
\pi^-)=2/3 R$ with $R=\Gamma(\pi\pi)/\Gamma(\pi\pi)+\Gamma(KK)\sim
0.68$ \cite {PDG} is used. It is clear that unless there exists
some other mechanism in the decay processes, it is difficult to
explain the large measured BR by just considering short-distant
interactions. Actually, the case is similar to the charmed decay
of $B^{+}\to \chi_{c0} K^{+}$ with its BR being
$(6.0^{+2.1}_{-1.8}\pm 1.1)\times 10^{-4}$ \cite{Belle3} and $(2.4
\pm 0.7)\times 10^{-4}$ \cite{Babar3} measured by Belle and BaBar,
respectively. In order to to agree with the experimental data, the
authors of Ref. \cite{CDP} suggest that the possible rescattering
effects, such as $B\to D^{*}_{s} D$, $B\to D_{s} D^{*}$ and $B\to
D^{*}_{s} D^{*}$ decays via triangle diagrams with the strong
couplings of $g_{DD\chi_{c0}}$, $g_{D^{*}D^{*}\chi_{c0}}$,
$g_{D^{*}D^{*}\chi_{c0}}$ etc., could enhance the BR and reach
$(1.1\sim 3.2)\times 10^{-4}$. We find that the mechanism could
also apply to $B\to f_{0}(980) K^{(*)}$ decays by replacing
$\chi_{c0}$ with $f_{0}(980)$ and taking the proper couplings such
as $g_{D^{*}_{s}D_{s} f_{0}(980)}$ etc.. The illustrated diagrams
are displayed in
Figure \ref{res}. 
Since the differences between $f_{0}(980) K^{(*)}$ and $
\chi_{c0}K^{+}$ modes are only related to the strong coupling
constants, by following the procedure of Ref. \cite{CDP} and
taking proper values of strong coupling constants, one expects
that the rescattering effects in Figure \ref {res} can reach
${\cal O}(10^{-5})$ easily. In addition, we note that, as the
possibility of gluonium state in $\eta'$ \cite{KS,Ali} was
proposed to explain the tendency of large BR for $\eta'$
production \cite{CLEO-BELLE}, it is also possible to understand
$B\to f_0(980) K$ decays by considering $g-g-f_{0}(980)$ coupling.



Finally, it is worth to mention that like the decays of $B\to
J/\Psi K_{S,L}$ and $B\to \phi K_{S,L}$, $B\to f_{0}(980) K_{S,L}$
can be another outstanding candidate for the observation of the
time-dependent CP asymmetry, defined by
\begin{eqnarray}
{\bar{\Gamma}(t)-\Gamma(t) \over
\bar{\Gamma}(t)+\Gamma(t)}&=&-A_{CP}^{dir}(B_{d}\to f) \cos(\Delta
M_{d} t) - A_{CP}^{mix}(B_{d}\to f) \sin(\Delta M_{d})
\end{eqnarray}
with
\begin{eqnarray}
A_{CP}^{dir}(B_{d}\to f)&=& {1-|\lambda|^2 \over 1+|\lambda|^2} \nonumber \\
 A_{CP}^{mix}(B_{d}\to f)&=&{2Im\lambda \over 1+|\lambda|^2},\nonumber \\
\lambda&=&\eta e^{-2i\phi_{1}}{A(\bar{B}_{d}\to f) \over
A(B_{d}\to f) },\label{tcp}
\end{eqnarray}
where $f$ expresses the  CP final state classified by $\eta=\pm
1$, $A_{CP}^{dir}\ (A_{CP}^{mix})$ denotes the direct
(mixing-induced) CP violation and $A(B(\bar{B}_{d})\to f)$ are the
$B(\bar{B})$ meson decay amplitudes and
$\phi_{1}=Arg(V^{*}_{td}V_{tb})$. From Eq. (\ref{tcp}), we know
that if no CP violating phase is involved in the decay amplitude,
the direct CP violating observable will vanish and $\lambda=\eta
e^{-2i\phi_{1}}$. Consequently, the mixing-induced CP asymmetry is
only related to $\sin2\phi_{1}$. Fortunately, since there is no
tree contribution, $B\to f_{0}(980) K_{S,L}$ decays satisfy the
criterion so that the mixing-induced CP asymmetry of $B\to
f_{0}(980) K_{S,L}$ is expected to be the same as that measured in
$B\to J/\Psi K_{S,L}$ decays, except a small derivation from
higher order contributions \cite{GW}.

We have studied $f_{0}(980)$ scalar meson production in B meson
decays by assuming that its flavor contents are $\bar{s} s$ and
$\bar{n} n$ states. We have found the role of $\bar{n} n$ on the
BRs of $B\to f_{0}(980) K$ is crucial. We have also pointed out
that nonvanishing BR of $B^{+}\to f_{0}(980) \pi^{+}$ could test
the existence of the $\bar{n} n$ content although the expected BR
is less than $10^{-6}$. If the BRs of $B\to f_{0}(980) K^{(*)}$
are measured to be $10^{-5}$ in experiments, the results could be
the evidence of existing rescattering effects in heavy $B$ meson
decays. On the other hand, although we concentrate on the $\bar{q}
q$ states of $f_{0}(980)$, we do not exclude the possibilities of
four-quark $qq\bar{q}\bar{q}$ and gluonium states. The
possibilities could be clarified by other experiments, such as
$\phi\to \pi\pi(KK)\gamma$ \cite{AG}, $D^{+}_{s}\to \pi^{+}\pi^{+}
\pi^{-}$ \cite{E791} decays and $B^{+}\to f_{0}(980) \pi^{+}$ with
the measured BR of ${\cal{O}}(10^{-6})$. Furthermore, the more
general approach to deal with the decays of $B\to f_{0}(980) K$
can refer to Ref. \cite{CLPRL}. We have also suggested that
time-dependent CP asymmetry in $B\to f_{0}(980) K_{S,L}$ could be
a new explorer of $\sin2\phi_{1}$. We note that our study of $B$
decays with $f_{0}(980)$ in the final state can be applied to
other modes with scalar or isoscalar final states such as
$\sigma(600)$, $a_{0}(980)$ etc.. Remarkably, $B$ factories have
opened a new opportunity to understand the contents of scalars and
their
productions.\\

\noindent {\bf Acknowledgments:}

The author would like to thank H.N. Li, Y.Y. Keum, H.Y. Cheng,
C.W. Hwang, K.C. Yang and C.Q. Geng for their useful discussions.
This work was supported in part by the National Science Council of
the Republic of China under Grant No. NSC-91-2112-M-001-053 and by
the National Center for Theoretical Science. And also, the idea is
created during attending Summer Institute 2002 at Fuji-Yoshida,
Japan.\\



\begin{thebibliography}{99}
\bibitem{Belle1} Belle Collaboration, A. Garmash {\it et al.}, Phys. Rev. D
{\bf 65}, 092005 (2002).
\bibitem{BaBar1} BaBar Collaboration, B. Aubert  {\it et al.}, hep-ex/0206004.

\bibitem{f0} S.D. Protopopescu {\it et al.}, Phys. Rev. D{\bf 7},
1279 (1973); B. Hyams {\it et al.}, Nucl. Phys. B{\bf 64}, 4
(1973).

\bibitem{4q} R.J. Jaffe, Phys. Rev. D{\bf 15}, 267 (1977); {\it
ibid}, 281 (1997); M. Alford and R.J. Jaffe, Nucl. Phys. B{\bf
578}, 367 (2000).

\bibitem{KK} J. Weinstein and N. Isgur, Phys. Rev. Lett. {\bf 48},
659 (1982); Phys. Rev. D{\bf 27}, 588 (1983); Phys. Rev. D{\bf
41}, 2236 (1990); M.P. Locher {\it etal.}, Eur. Phys. J. C{\bf 4}
317 (1998).

\bibitem{qq} N.A. Tornqvist, Phys. Rev. Lett. {\bf 49}, 624
(1982); N.A. Tornqvist and M. Roos, Phys. Rev. Lett. {\bf 76},
1575 (1996).

\bibitem{DP} F. De Fazio and M.R. Pennington, Phys.
Lett. B{\bf 521}, 15 (2001).

\bibitem{DLS} R. Delbourgo, D. Liu, and M.D. Scadron, Phys. Lett.
B{\bf 446}, 332 (1999); T.M. Aliev {\it et al.}, Phys. Lett. B{\bf
527}, 193 (2002).

\bibitem{AAN} A.V. Anisovich, V.V. Anisovich and V.A. Nikonov,
hep-ph/0011191.

\bibitem{KVRS} F. Kleefeld {\it et al.}, Phys. Rev. D{\bf 66},
034007 (2002).

\bibitem{dfk} E. van Beveren, G. Rupp and M.D. Scadron, Phys.
Lett. B{\bf 495}, 300 (2000); A. Deandrea {\it et al.}, Phys.
Lett. B{\bf 502}, 79 (2001).

\bibitem{BBL}  G. Buchalla, A.J. Buras and M.E. Lautenbacher, Rev. Mod.
Phys. {\bf 68}, 1230 (1996).

\bibitem{Chen-PLB525} C.H. Chen, Phys. Lett. B{\bf 525}, 56
(2002).

\bibitem{CKL-PRD} C.H. Chen, Y.Y. Keum and H.N. Li, Phys. Rev.
D{\bf 64}, 112002 (2001); S. Mishima, Phys. Lett. B{\bf 521}, 252
(2001).

\bibitem{Belle2} Belle Collaboration, K.F. Chen, talk given
at ICHEP2002 hold on 24 - 31 July 2002, Amsterdam, Netherlands.

\bibitem{BaBar2} BaBar Collaboration, A. Telnov, talk given at APS's
2002 meeting of the division of particles and fields hold on 24 -
28 May 2002, Williamsburg, Virginia, USA.


\bibitem{KLS-PRD} T. Kurimoto, H.N. Li, A.I. Sanda Phys. Rev. D{\bf 65}
014007 (2002).

\bibitem{BBKT} P. Ball, JHEP 9809, 005 (1998); P. Ball {\it et. al.},
Nucl. Phys. B{\bf 529}, 323 (1998).


\bibitem{KLS}  Y.Y. Keum, H.N. Li, and A.I. Sanda, Phys. Lett. B{\bf 504},
6 (2001); Phys. Rev. D{\bf 63}, 054008 (2001).

\bibitem{LUY}  C.D. L${\rm \ddot{u}}$, K. Ukai, and M.Z. Yang, Phys. Rev. D%
{\bf 63}, 074009 (2001).

\bibitem{CL-PRD} C.H. Chen and H.N. Li, Phys. Rev. D{\bf 63}, 014003 (2001).

\bibitem{Chen-PLB520}C.H. Chen, Phys. Lett. B{\bf 520}, 33 (2001).

\bibitem{Melic} B. Melic, Phys. Rev. D{\bf 59}, 074005 (1999).

\bibitem{KS} E. Kou and A.I. Sanda, Phys. Lett. B{\bf 525}, 240 (2002).

\bibitem{SU} A.I. Sanda and K. Ukai, Prog. Theor. Phys. {\bf 107}, 421 (2002).

\bibitem{Keum} Y.Y. Keum, H-n. Li, and A.I. Sanda, hep-ph/0201103;
Y.Y. Keum, hep-ph/0209002; hep-ph/0209014.

\bibitem{CQ} C.H. Chen and C.Q. Geng, Nucl. Phys. B{\bf 636}, 338
(2002).


\bibitem{Ochs} P. Minkowski and W. Ochs, hep-ph/0209223; W. Ochs,
hep-ph/0111309; C.M. Shakin, Phys. Rev. D{\bf 65}, 114011 (2002).

\bibitem{PDG} K. Hagiwara {\it et al.}, Phys. Rev. D{\bf 66}, 010001 (2002).

\bibitem{Belle3} Belle Collaboration, Phys. Rev. Lett. {\bf 88},
031802 (2002).

\bibitem{Babar3}BaBar Collaboration, hep-ex/0207066.

\bibitem{CDP} P. Colangelo, F. De Fazio and T.N. Pham, Phys. Lett.
B{\bf 542}, 71 (2002).

\bibitem{Ali} D. Atwood and A. Soni, Phys. Rev. Lett. {\bf 79}, 5206 (1997);
A. Ali and A.Y. Parkhomenko, Phys. Rev. D{\bf 65}, 074020 (2000).

\bibitem{CLEO-BELLE} CLEO Collaboration, T.E. Browder {\it et al.}, Phys. Rev.
Lett. {\bf 81}, 1786 (1998); Belle Collaboration, K. Abe {\it et
al.}, Phys. Lett. B{\bf 517}, 309 (2001).

\bibitem{GW} Y. Grossman and M.P. Worah, Phys. Lett. B{\bf 395}, 241 (1997).

\bibitem{AG} N.N. Achasov and V.V. Gubin, Phys. Rev. D{\bf 64},
094016 (2001);  A. Antonelli, invited talk at the XXII Physics in
Collisions Conference (PIC02), Stanford, Ca, USA, June 2002,
hep-ex/0209069.

\bibitem{E791} E791 Collaboration, E.M. Aitala {\it et al.}, Phys.
Rev. Lett. {\bf 86}, 765 (2001).

\bibitem{CLPRL} C.H. Chen and H.N. Li, hep-ph/0209043.

\end{thebibliography}
\end{document}